\documentclass[12]{article}

\usepackage[utf8]{inputenc}
\usepackage[OT2,T1]{fontenc}
\DeclareSymbolFont{cyrletters}{OT2}{wncyr}{m}{n}
\DeclareMathSymbol{\ssha}{\mathalpha}{cyrletters}{"78}

\newcommand{\sha} { {\ssha}}

\usepackage{amssymb, amsmath, amscd}
\usepackage{graphicx} 
\usepackage{color}
\usepackage{subfig, color}
\usepackage{float}

\usepackage{soul}

\newtheorem{definition}{Definition}

\newtheorem{proposition}{Proposition}

\numberwithin{equation}{section}
\newtheorem{example}{Example}
\usepackage{tikz-cd}
\usepackage{rotating}

\begin{document}


\newcommand{\calA}{{\cal A}}
\newcommand{\calB}{{\cal B}}
\newcommand{\calF}{{\cal F}}
\newcommand{\calG}{{\cal G}}
\newcommand{\calR}{{\cal R}}
\newcommand{\calZ}{{\cal Z}}
\def\D{\mathcal{D}}
\def\L{\mathcal{L}}
\def\S{\mathcal{S}}
\def\I{\mathcal{I}}
\def\V{\mathcal{V}}
\def\E{\mathcal{E}}
\def\M{\mathcal{M}}

\def\A{\mathscr{A}}
\def\F{\mathscr{F}}
\def\G{\mathscr{G}}

\newcommand{\N}{{\mathbb N}}
\newcommand{\Z}{{\mathbb Z}}
\newcommand{\Q}{{\mathbb Q}}
\newcommand{\R}{{\mathbb R}}
\newcommand{\CC}{{\mathbb C}}

\newcommand{\K}{{\mathbb K}}
\newcommand{\kk}{{\mathrm k}}

\def\J{\mathrm{J}}
\def\catC{{\bf \mathrm{C}}}
\def\x{\mathrm{x}}
\def\a{\mathrm{a}}
\def\d{\mathrm{d}}
\def\Bi{\mathrm{Bi}}
\def\op{\mathrm{op}}
\def\res{\mathrm{res}}
\def\span{\mathrm{span}}

\newcommand{\C}{{\bf C}}
\newcommand{\Objects}{{\bf Objects}}
\newcommand{\Arrows}{{\bf Arrows}}
\newcommand{\Sets}{{\bf Sets}}

\def\2F1{\mbox{ $_2${F}$_1$}}
\def\1F1{\mbox{ $_1${F}$_1$}}
\def\1F2{\mbox{ $_1${F}$_2$}}
\def\0F1{\mbox{ $_0${F}$_1$}}

\def\GL{\mathrm{GL}}
\def\det{\mathrm{det}}
\def\SL{\mathrm{SL}}
\def\PSL{\mathrm{PSL}}
\def\PGL{\mathrm{PGL}}
\def\O{\mathrm{O}}

\def\gl{\mathfrak{gl}}
\def\g{\mathfrak{g}}
\def\h{\mathfrak{h}}
\def\frakM{\mathfrak{M}}

\newcommand{\Frac}[2]{\displaystyle \frac{#1}{#2}}
\newcommand{\Sum}[2]{\displaystyle{\sum_{#1}^{#2}}}
\newcommand{\Prod}[2]{\displaystyle{\prod_{#1}^{#2}}}
\newcommand{\Int}[2]{\displaystyle{\int_{#1}^{#2}}}
\newcommand{\Lim}[1]{\displaystyle{\lim_{#1}\ }}

\newenvironment{menumerate}{%
    \renewcommand{\theenumi}{\roman{enumi}}%
    \renewcommand{\labelenumi}{\rm(\theenumi)}%
    \begin{enumerate}} {\end{enumerate}}

\newenvironment{system}[1][]%
	{\begin{eqnarray} #1 \left\{ \begin{array}{lll}}%
	{\end{array} \right. \end{eqnarray}}

\newenvironment{meqnarray}%
	{\begin{eqnarray}  \begin{array}{rcl}}%
	{\end{array}  \end{eqnarray}}

\newenvironment{marray}%
	{\\ \begin{tabular}{ll}}
	{\end{tabular}\\}

\newenvironment{program}[1]%
	{\begin{center} \hrulefill \quad {\sf #1} \quad \hrulefill \\[8pt]
		\begin{minipage}{0.90\linewidth}}
	{\end{minipage} \end{center} \hrule \vspace{2pt} \hrule}

\newcommand{\entrylabel}[1]{\mbox{\textsf{#1:}}\hfil}
\newenvironment{entry}
   {\begin{list}{}%
   	{\renewcommand{\makelabel}{\entrylabel}%
   	  \setlength{\labelwidth}{40pt}%
   	  \setlength{\leftmargin}{\labelwidth + \labelsep}%
   	}%
   }%
   {\end{list}}

\newenvironment{remark}{\par \noindent {\bf Remark. }}
			{\hfill $\blacksquare$ \par}

\newenvironment{Pmatrix}
        {$ \left( \!\! \begin{array}{rr} }
        {\end{array} \!\! \right) $}

\newcommand{\fleche}[1]{\stackrel{#1}\longrightarrow}
\def\ssi{si et seulement si\ }
\newcommand{\tab}{\hspace*{\fill}}
\newcommand{\bs}{{\backslash}}
\newcommand{\eps}{{\varepsilon}}
\newcommand{\into}{{\;\rightarrow\;}}
\newcommand{\PD}[2]{\frac{\partial #1}{\partial #2}}
\def\Hat{\widehat}
\def\Bar{\overline}
\def\vect{\vec}
\def\fbar{{\bar f}}
\def\xbar{{\bar \x}}
\newcommand{\afaire}[1]{$$\vdots$$ \begin{center} {\sc #1} \end{center} $$\vdots$$ }
\newcommand{\pref}[1]{(\ref{#1})}

\def\Maple{{\sc Maple}}
\def\RG{{\sc Rosenfeld-Gr\"obner}}



\newcommand{\algf}{\sffamily}
\newcommand{\BEGIN}{{\algf begin}}
\newcommand{\END}{{\algf end}}
\newcommand{\IF}{{\algf if}}
\newcommand{\THEN}{{\algf then}}
\newcommand{\ELSE}{{\algf else}}
\newcommand{\ELIF}{{\algf elif}}
\newcommand{\FI}{{\algf fi}}
\newcommand{\WHILE}{{\algf while}}
\newcommand{\FOR}{{\algf for}}
\newcommand{\DO}{{\algf do}}
\newcommand{\OD}{{\algf od}}
\newcommand{\RETURN}{{\algf return}}
\newcommand{\PROCEDURE}{{\algf procedure}}
\newcommand{\FUNCTION}{{\algf function}}
\newcommand{\INDENTER}{{\algf si} \=\+\kill}

\newcommand{\target}{\mathop{\mathrm{t}}}
\newcommand{\source}{\mathop{\mathrm{s}}}
\newcommand{\trdeg}{\mathop{\mathrm{tr~deg}}}
\newcommand{\jet}[2]{\jmath_{#1}^{#2}}
\newcommand{\rank}{\operatorname{rank}}
\newcommand{\sign}{\operatorname{sign}}
\newcommand{\ord}{\operatorname{ord}}
\newcommand{\aut}{\operatorname{aut}}
\newcommand{\Hom}{\operatorname{Hom}}
\newcommand{\myhom}{\operatorname{hom}}
\newcommand{\codim}{\operatorname{codim}}
\newcommand{\coker}{\operatorname{coker}}
\newcommand{\rp}{\operatorname{rp}}
\newcommand{\leader}{\operatorname{ld}}
\newcommand{\card}{\operatorname{card}}
\newcommand{\Fr}{\operatorname{Frac}}
\newcommand{\RF}{\operatorname{\mathsf{reduced\_form}}}
\newcommand{\rang}{\operatorname{rang}}

\def \Id{\mathrm{Id}}

\def \diff{\mathrm{Diff}^{\mathrm{loc}} }
\def \diffg{\mathrm{Diff} }
\def \Esc{\mathrm{Esc}}

\newcommand{\initial}{\mathop{\mathsf{init}}}
\newcommand{\separant}{\mathop{\mathsf{sep}}}
\newcommand{\quo}{\mathop{\mathsf{quo}}}
\newcommand{\pquo}{\mathop{\mathsf{pquo}}}
\newcommand{\lcoeff}{\mathop{\mathsf{lcoeff}}}
\newcommand{\mvar}{\mathop{\mathsf{mvar}}}

\newcommand{\prem}{\mathop{\mathsf{prem}}}
\newcommand{\remp}{\mathrel{\mathsf{partial\_rem}}}
\newcommand{\remf}{\mathrel{\mathsf{full\_rem}}}
\renewcommand{\gcd}{\mathop{\mathrm{gcd}}}
\newcommand{\pairs}{\mathop{\mathrm{pairs}}}
\newcommand{\dd}{\mathrm{d}}
\newcommand{\ideal}[1]{(#1)}
\newcommand{\cont}{\mathop{\mathrm{cont}}}
\newcommand{\pp}{\mathop{\mathrm{pp}}}
\newcommand{\pgcd}{\mathop{\mathrm{pgcd}}}
\newcommand{\ppmc}{\mathop{\mathrm{ppcm}}}
\newcommand{\init}{\mathop{\mathrm{initial}}}

\title{
{Knuth-Bendix Completion Algorithm and Shuffle Algebras
For 
Compiling NISQ  Circuits}}

\author{Raouf Dridi\footnote{rdridi@andrew.cmu.edu}, \, Hedayat Alghassi\footnote{halghassi@cmu.edu} ,\, Sridhar Tayur\footnote{stayur@cmu.edu} \\
~~\\
\small  { Quantum Computing Group}\\
\small {Tepper School of Business}\\
\small  {{ Carnegie Mellon University, Pittsburgh, PA 15213}}\\}

\maketitle

\begin{abstract}
    Compiling quantum circuits
    lends itself to an elegant formulation in the language of   rewriting systems  on non commutative polynomial algebras   $\mathbb Q\langle X\rangle$. The alphabet $X$ is the set of the allowed hardware 2-qubit gates.  The set of gates that we wish to implement from $X$ are elements of a free monoid $X^*$ (obtained by concatenating the letters of $X$). In this setting, compiling an idealized gate is equivalent to computing its unique normal form with respect to the rewriting system $\mathcal R\subset \mathbb Q\langle X\rangle$ that encodes the hardware constraints and capabilities. This system $\mathcal R$ is generated using two different mechanisms: 1) using the Knuth-Bendix completion algorithm on the algebra $\mathbb Q\langle X\rangle$, and 2) using the Buchberger algorithm on the shuffle algebra $\mathbb Q[L]$ where $L$ is the set of Lyndon words on $X$. 
\end{abstract}

~~

\newpage

\tableofcontents

\section{Introduction}

 {\it Rewriting systems} (RS) are a natural choice as an algorithmic framework for quantum circuits compilation (QCC) on Noisy Intermediate Scale Quantum (NISQ) computers. NISQ computers are viewed as a first step towards fault-tolerance machines (\cite{preskill}). Skepticism remains as to whether quantum computing can ever be realized in practice, although enthusiasts view it as a holy grail of 21st century computing (and it is one of the U.S. National Science Foundation's ``Ten Big Ideas''). It is a well known fact that current hardware prototypes of gates are not even close to being useful. Nevertheless, considerable interest exists in understanding what one can do with what is now available and what can be reasonably expected in the near future. Consequently, building a robust mathematical foundation and algorithmic framework to study NISQ computing is an exciting research opportunity. This is the viewpoint that we take.
 
 ~~\\
 Compiling idealized logical quantum circuits onto early hardware prototypes (such as those from IBM, Google, and Rigetti) means mapping the logical gates of an idealized algorithm onto  actual hardware that has {\it geometric} (connectivity between qubits is limited) and {\it cross-talk} (limitations on parallel execution on gates that have close proximity) constraints. Additionally, different gate operations take different times, so optimizing how to place a subset of them in parallel (whenever possible) becomes an important task.  Noise is not directly modeled; instead, the goal is to minimize the {\it makespan} of the proposed solution, with the belief that the {\it duration minimization} of the circuit to be executed is a good proxy to achieve the lowest decoherence. Our RS approach is an alternative to the use of constraint programming (CP) and artificial intelligence (AI) methods that have been explored in \cite{venturelli2019}. 

~~\\
 The theory of rewriting systems is well known to mathematicians and theoretical computer scientists. It is used to provide concrete and effective equality testing criteria between ideals of various algebras. It was initiated by 
D. Knuth and P. Bendix in \cite{Knuth} and 
by B. Buchberger \cite{Buchberger} in the commutative case, both in the 1970s. In our context of compilation into realistic quantum hardware architectures (such as those from IBM, Google, and Rigetti), 
{\it compiling an idealized gate $w$ consists of computing its unique normal form  with respect to a rewriting system $\mathcal{R}$} that is  computed from the initial set of hardware constraints, 
using the Knuth-Bendix algorithm on non commutative algebras, or, alternatively, with the Buchberger algorithm on commutative algebras with Radford bases. Table 1 provides the dictionary between gate circuit compiling and rewriting systems. Our work here, mathematically speaking, is an extension of the algebraic geometry ideas of \cite{cmu1} to the non commutative case, which  were developed for compiling on hardware based on adiabatic quantum computing (AQC), such as that from D-Wave, which is an entirely different paradigm of quantum computing. 


\begin{table}[h]
\centering
\begin{tabular}{|c|c|}
\hline\\
Compiling quantum circuits & Rewriting systems \\
\hline \hline\\
Hardware gates & Alphabet $X$\\
\hline\\
Hardware constraints & A noetherian rewriting system $\mathcal{R}$\\\hline\\
Composition of gates &  Concatenation \\
\hline\\
Composite gate &  A word $w\in X^*$  \\
\hline\\
Compiling a gate $w$ & Computing the normal form ${\sf NF}_\mathcal R(w)$
\\\hline\\
Clock cycles & Monomial order\\\hline
\hline\\
Parallel gates & Commuting words\\\hline
\end{tabular}
\caption{Correspondence between gate circuit compiling and rewriting systems.}
\end{table}

\subsection{Illustrative Examples}
Two examples introduce the concept of compiling gates into restricted architectures (and   reveal the basic intuition of) using rewriting systems.

\begin{example}
Figure \ref{ibm} (right) shows the connectivity  graph of IBM QPU ibmqx2~\cite{ibm}. The directions in the graph restricts (the already restricted) set of controlled not (cnot) gates that can be directly represented. The edge $0\rightarrow 1$ corresponds to the cnot gate $cnot_{01}$ with control qubit at node 0 and target qubit at node 1.

~~\\
The gate $cnot_{10}$ can not be directly represented. The remedy in this simple example is easy: swap the two qubits 0 and 1, then apply $cnot_{01}$: we can write   
$cnot_{10} = s_{01}cnot_{01}$. 
	 \begin{figure}[H]
    	\centering
    		\subfloat  
    		{{\includegraphics[scale=.3]{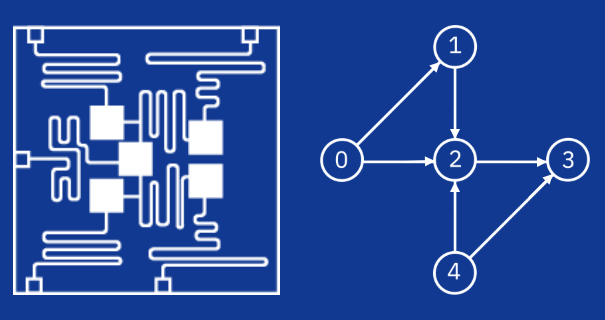} }}%
    		\caption{{IBM QPU (ibmqx2): Left) Quantum chip structure, consists of five superconducting transmon qubit, kept at $0.015$ Kelvin, built from Niobium capacitors and Josephson junctions and   microwave resonators for inter-qubit connections. Right) Graph representation, consists of five nodes and six directional edges.}}
           \label{ibm}
 	 \end{figure}%
~~\\
In the language of the theory of rewriting systems, the set of the hardware (also called primitive) gates 
\begin{equation}
X= \{cnot_{01}, \, cnot_{02},\, cnot_{12},\cdots \} \cup \{s_{01},  s_{02}, s_{12},\cdots \}    
\end{equation}
 constitutes an alphabet:
 \begin{itemize}
     \item Any concatenation of letters of $X$ is word, and we write $X^*$ for the set of all words generated by $X$. We have (for instance)
$cnot_{10}\in X^*.$ 
\item We also define on $X^*$ an ordering relation that reflects the total runtime of each word. As an example, we have $ s_{01}cnot_{01} \succ cnot_{01}$. 
\item The particular topology of chip gives rise to a number of  relations between the letters of $X$ such as the relation $s_{01}^2=Identity$. We represent this relation  as a substitution (replacement) rule: 
\begin{equation}
s_{01}s_{01}\rightarrow \varepsilon    
\end{equation}
where  $\varepsilon$ is the empty word. Another rule is
\begin{equation}
(s_{01}s_{12})^3\rightarrow\varepsilon.    
\end{equation}
In general, a rule replaces a word with an expression  of smaller value with respect to the ordering relation $\succ$. In our examples, the two words on the right hand side of the rules reduce to the empty word.  
\item Because different gates can have different runtimes, the alphabet is ordered accordingly.
\end{itemize}
~~\\
The set of rules comes together to form a rewriting system that represents the given chip. This rewriting system is constructed only once (per hardware architecture).  
\end{example}

\begin{example}
 
	 \begin{figure}[H]
    	\centering
    		\subfloat  
    		{{\includegraphics[scale=0.6]{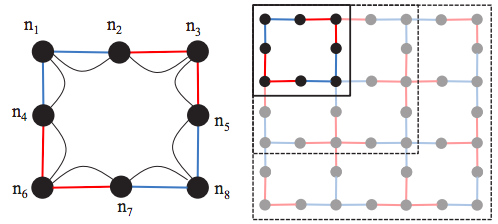} }}%
    		\caption{{ Hypothetical chip layout, based on an architecture proposed by Rigetti Computing \cite{Davide2018}.    }}
           \label{davide0}
 	 \end{figure}%

Consider Figure \ref{davide0}  from \cite{Davide2018}.   Three types of ``2-qubit gates''  are built-in: Swap gates, represented by the black curved lines between the nodes and denoted by $s_{ij}$, and the red gates $r_{ij}$ and blue gates $b_{ij}$, which are two mixing gates in QAOA (used in \cite{Davide2018} for the MaxCut problem).  Here, it suffices to consider them as specific gates operating as indicated (without the need to know their explicit actions). The collection of all of these gates, which
have different runtimes, form the alphabet. This alphabet is ordered according to the execution duration of the gates. 

~\\
The alphabet $X$ is given by the set of swaps $s_{ij}$, the red and blue gates $r_{ij}, \, b_{ij}$ for the allowed $i$ and $j$, as indicated in Figure~\ref{davide0}.  The sequence of swaps $s_{12}, \, s_{23}$ is a word $w=s_{12}s_{23}$. Another word is
$s_{14}s_{23}s_{12}r_{2,3}$, which implements the gate ${\sf Red}_{2, 4}$. Note that the symbol ${\sf Red}_{2, 4}$ is not part of the algebra~$\mathbb Q\langle X\rangle$ generated by $X$ (see next section). 
\end{example}

~~\\
The paper is structured as follows: In Section 2 we recast the problem of compiling quantum circuits in the language of non commutative algebras. We explore different tools that come with such algebras. In Section 3, we define monomial orders used in the construction of rewriting systems as well as in the computation of normal form.  Section 4 reviews the Knuth-Bendix completion algorithm and some basic background in the theory of rewriting systems. We explain the compiling process in this language. Section 5 discusses parallelism, encoding of cross-talk constraints, and initial qubit placement. Section 6   presents an alternative approach to the  Knuth-Bendix algorithm through the use of shuffle algebras. We conclude in Section 7.

\section{Problem formulation}
We write $X$ for the finite set of the hardware 2-qubits gates. In reality, to each hardware gate, we assign a {\it letter} $x\in X$. We will capture/encode the particular properties (as well as restrictions) of these gates, as we move forward through algebraic relations between their corresponding letters. We also write $X^*$ for the free monoid generated by the set~$X$. A {\it word} in $X^*$ is a concatenation of letters of $X$.  The length of each word is the number of its letters. The {\it empty} word is denoted by $\varepsilon$. We write $X^+=X^* - \{\varepsilon\}$. 

~~\\
We consider the free associate algebra 
$\mathbb{Q}\langle X \rangle$ of the polynomials, with rational coefficients and non commutative variables $x\in X$. That is, $\mathbb{Q}\langle X \rangle$ is given by all  $\mathbb{Q}-$linear combinations of words.  The free Lie algebra, denoted  $\mathcal{L}_\mathbb{Q}\langle X\rangle$ or simply
$\mathcal{L}\langle X\rangle$, is defined as the quotient:
\begin{equation}
 \mathcal{L}\langle X\rangle = \mathbb{Q}\langle X \rangle/ \mathcal{J}    
\end{equation}
where $\mathcal{J}$ is the ideal generated by 
\begin{system}
 Q(x) &=& [x,x]\\
 J(x, y,z) &=&  [x, [y, z]] +  [y, [z, x]] +  [z, [x, y]]
\end{system}%
for $x, y$ and $z$ in $\mathbb{Q}\langle X \rangle$. Note that for every $x$, the map $ad_x$ acts as a derivation (with $ad_x$ being the adjoint representation of $x$). 
The Lie algebra $\mathcal{L}\langle X\rangle$ is the place where the structural particularities (e.g., constraints) of  the given chip, as well as the   particularities  of its set of allowed primitive gates, are expressed as algebraic relations between the letters of the alphabet $X$.  The ensemble of these relations will constitute   
our  rewriting system   $\mathcal{R}$. The compiling problem translates then into computing the unique normal forms (finite sequence of substitutions) with respect to the rewriting system $\mathcal R$. Each normal form is the {\it minimal}  representative  of the equivalence class of the gate $w$, which we wish to compile, with respect to the following relation ({\it Church-Roser property}):
\begin{equation}
    w_1 \sim_\mathcal R w_2 \Leftrightarrow {\sf NF}_\mathcal R(w_1) = {\sf NF}_\mathcal R(w_2).
\end{equation}
The next section shows how this notion of
 minimality (of the normal form) coincides with what one desires in this application: \textit{consuming the fewest resources}.

\section{Number of clock cycles and monomial order}
Let $\mathbb{Q}\langle X \rangle$ be the non commutative algebra as above. 
The following definition is very similar to the commutative case (which can be found in \cite{cox} for instance). 
\begin{definition}[Monomial order]
A monomial order on $\mathbb{Q}\langle X \rangle$ is any relation~$\succ$ on the set of words $X^*$ satisfying
\begin{itemize}
    \item[(a)] $\succ$ is a total ordering relation;
    \item[(b)] $\succ$ is compatible with multiplication in $\mathbb{Q}\langle X \rangle$;
    \item[(c)] $\succ$ is a well-ordering. That is, every nonempty collection of monomials has a smallest element under $\succ$.
\end{itemize}
\end{definition}
Condition (c) ensures that processes that work on collections of monomials, for example, the collection of all monomials less than some fixed monomial $w$, will terminate in a finite number of steps (see Noetherian systems below).

~~\\
A mononial order $\succ$ can always be represented by a matrix $M$. If ${\bf w}_1, \cdots, {\bf w}_m$ are the rows of $M$, and $x^\alpha, x^\beta\in X^*$, then
$x^\alpha \succ x^\beta$ if there is an $\ell \leq m$ such that $\alpha\cdot {\bf w}_i =\beta \cdot {\bf w}_i$ for $i=1,\cdots, \ell-1,$ but $\alpha \cdot {\bf w}_\ell >\beta \cdot {\bf w}_\ell$. 
%

~~\\
There are many choices of monomial orders on $\mathbb{Q}\langle X\rangle$. In our case, we need to take into account the different duration of the gates/letters $x\in X$. 
\begin{example}\label{runningexample}
We continue with our example above of Rigetti architecture. The duration of a swap gate is $\tau_{s}=2$ (clock cycles) whilst the duration of  blue and red gates  are $\tau_{blue}=3$ and $\tau_{red}=4$, respectively. The row ${\bf w} := {\bf w}_1$ is then given by
the row vector
\begin{equation}
    {\bf w} = (2/4, \, 3/4, \, 1)
\end{equation}
For instance, the word $s_{1,2}$ is less than the word $b_{1,2}$ because 
\begin{equation}
(1, 0,0)\cdot  {\bf w}=1/2 <  (0, 1,0)\cdot  {\bf w}= 3/4.    
\end{equation}
 However, the word $s_{1,2} s_{2,3}$ is bigger than $b_{1,2}$ because now 
 \begin{equation}
 (2, 0,0)\cdot  {\bf w}=1 > (0, 1,0)\cdot  {\bf w}= 3/4.    
 \end{equation}
  We require that if the set of letters of a given word commute then this word has the duration of its largest letter (see Section \ref{parallel}, which discusses parallelism).
\end{example}

~~\\
We end this section with the following definitions:
\begin{itemize}
    \item The largest word of a polynomial $p\in \mathbb{Q}\langle X \rangle$ is called the {\it leading term} and denoted by $lt(p)$.
    \item The coefficient of this word is called the {\it leading coefficient} of $p$ and denoted by $lc(p$).
    \item  The polynomial $p$ deprived of its leading term is denoted by $rest(p)$.
\end{itemize}
 As a matter of fact, a rewriting system is a set of rules $w\rightarrow p$ such that $w\succ lt(p)$, plus other properties, which we describe next.

\section{Rewriting systems and the Knuth-Bendix completion algorithm}
The technology of rewriting systems has been used extensively to obtain effective equality criteria  between vector spaces (respectively, ideals) of polynomials defined by a finite number of generators. The Buchberger algorithm \cite{Buchberger, cox} is an example for ideal of commutative polynomials.   Here, we review this concept in its most relaxed formulation (generalized algebra \cite{Knuth}), valid for commutative and non commutative polynomials.  

\begin{definition}
A {\it rewriting system} $\mathcal R$ on $\mathbb{Q}\langle X\rangle$   is a finite set of {\it rules} of the form 
\begin{equation}
    w_i\rightarrow q_i
\end{equation}
where $w$ is a word and $q_i$ is a polynomial. The word $w$ is required to be larger than words of $q_i$.  In other words, we consider a set of polynomials
\begin{equation}
    G = \{g_i = w_i-q_i \, |\quad lt(g_i)=w_i, 1\leq i\leq n\}.
\end{equation}
When a polynomial $q$ is deduced from $p$, by applying a number of rules, we write~{$p^{\, \underrightarrow{*}\, }q $}.
\end{definition} 
~~\\
A polynomial is said to be {\it reduced} by the system $\mathcal{R}$ if there are no more rules that can be applied. A {\it normal form} of the   polynomial $p$  is a reduced polynomial $q$ such that $p^{\, \underrightarrow{*}\, }q $.

\begin{definition}[Noetherian and confluent systems]
~\\
\begin{itemize}
    \item If $p \succ q$ for every $p$ and $q$ such that $p^{\, \underrightarrow{*}\, }q $ then the rewriting system $\mathcal{R}$ is {\it noetherian}.
    \item The rewriting system $\mathcal R$ is said to be {\it confluent} if and only if
$p^{\, \underrightarrow{*}\, }q_1 $ 
and
$p^{\, \underrightarrow{*}\, }q_2$ implies
the existence of $q$ such that
$q_1^{\, \underrightarrow{*}\, }q $
and
$q_2^{\, \underrightarrow{*}\, }q $.
\end{itemize}

\end{definition}

\begin{example}
Examples 1-3 are noetherien (by the condition (c) in the definition of monomial orders) and also confluent.
\end{example}

~~\\
A {\it critical pair} is a pair of rules of the form
\begin{system}\label{cpair}
 uc &\rightarrow_{r_1}& p\\
 cv &\rightarrow_{r_2}& q\
\end{system}%
with $c \neq \varepsilon,  u, v\in X$ and $p, q\in \mathbb{Q}\langle X\rangle$. If a such pair exists, the word $w=ucv$ can be reduced into two polynomials $pv$ and $uq$. The polynomial $s=pv-uq$ is called an {\it S--polynomial}. It suffices for a system to be confluent on its critical pairs to be confluent. 

\begin{proposition}
In a noetherian and confluent system, every polynomial has a unique normal form.
\end{proposition}

~~\\
The Knuth-Bendix completion algorithm  completes a non confluent system into a confluent one by adding new rewrite rules (S--polynomials). It works as follows: 

\begin{itemize}
    \item Input: List of polynomials $G =\{g_i=w_i-q_i\}$.
    \item Ouput: Confluent system $\mathcal{R}$.
    \item Initialize $\mathcal{R}=G$.
    \item {\sf WHILE} $G$ has critical pairs {\sf DO}
    \begin{itemize}
        \item reduce the critical pair and complete $G$ with the normal forms of the obtained S--polynomials.
        \item Compute the new critical pairs.
    \end{itemize}
    \item {\sf RETURN} $\mathcal R$
\end{itemize}


\begin{example}
Continuing  with Example \ref{runningexample},  
we would like to compile the gate ${\sf Red}_{2, 4}$. We can do that with the word
\begin{equation}\label{w1}
    w_1 =   s_{14}s_{12}s_{23}s_{12}r_{23}.
\end{equation}
However, there is a better sequence, which uses fewer swap gates:
\begin{equation}
w_2 := s_{14}s_{23}s_{12}r_{23} .    
\end{equation}
The question is how our approach determines $w_2$.  First, recall that the alphabet $X$ is given by all swaps, blue and red gates, and ordered by the monomial ordering   given by
\begin{equation}
    {\bf w} =  \left (\frac{2}{4}, \, \frac{3}{4},\, 1\right).
\end{equation}
The rewriting system is given   by the hardware constraints that  contain, for instance, the three simple rules
(describing the order of permutation cycles)
\begin{system}\label{rules}
 {(s_{12}s_{23} )}^3&\rightarrow& \varepsilon,\\[3mm]
 {s_{12}}^2&\rightarrow& \varepsilon,\\[3mm]
 {s_{23}}^2 &\rightarrow& \varepsilon
\end{system}%
The first two rules are trivially a critical pair (with $c=s_{12}$ and $p=q=\varepsilon$ in the system (\ref{cpair})), because they expand into
\begin{system}
 s_{12}(s_{23} s_{12} s_{23} s_{12} s_{23})  &\rightarrow& \varepsilon\\[3mm]
  (s_{12})s_{12} &\rightarrow& \varepsilon.
\end{system}%
We have the S-polynomial 
\begin{equation}
   s_{23} s_{12} s_{23} s_{12} s_{23} - s_{12}, 
\end{equation}
which, in turns, gives the rule
\begin{equation}\label{favouriteRule}
   s_{23} s_{12} s_{23} s_{12} s_{23} \rightarrow s_{12}. 
\end{equation}
Continuing with the third rule of the system (\ref{rules}), the same calculation gives  
\begin{equation}
    s_{23} s_{12} s_{23} s_{12} \rightarrow s_{12}s_{23}.
\end{equation}
And finally, 
\begin{equation}
     s_{12} s_{23} s_{12} \rightarrow s_{23}s_{12}s_{23}.
\end{equation}
Another  hardware rule is relevant here: ${s_{23}r_{23}  \rightarrow r_{23}}$, expressing the fact that the gate $r_{23}$ is undirected. Thus, we have the two rules: 
\begin{system}
 s_{12} s_{23} s_{12} &\rightarrow& s_{23}s_{12}s_{23}\\[3mm]
 s_{23} r_{23} &\rightarrow& r_{23}
\end{system}%
The normal form  of  $w_1$ with respect to these two rules gives  $w_2$. 

~~\\
Suppose now  we want to compile the gate
${\sf Red_{2,4}\, \&\&\,  Blue_{1,2}}$ where ${\sf \&\&}$ is the logical conjunction (and) operator. Consider  two implementations:
\begin{equation}\label{w3}
  w_3 := b_{12} s_{14; 23} s_{12} r_{2,3}
\end{equation}
and
\begin{equation}
     w_4 := s_{14; 23} s_{12} s_{14; 23}  r_{23}b_{12}.
\end{equation}
The gate $s_{14; 23}$ indicates that
the two swaps are run in parallel. 
We have ${\bf w}\cdot w_3 = {\bf w} \cdot (b + 2 s +r )= 11/4$ while ${\bf w} \cdot w_4 = {\bf w} \cdot (3s +r +b) =13/4$. So the algorithm outputs $w_3$. 
\end{example}
The symmetrical structure of the connectivity graphs can be used to avoid redundant rules and redundant critical pairs calculations. Similar ideas were used in \cite{cmu1} in the context of graph minor embedding that increases the speed of computation. 

\section{Parallelism, cross-talk, and initial  placement}\label{parallel}
\begin{figure}[H]
    	\centering
    		\subfloat  
    		{{\includegraphics[scale=0.6]{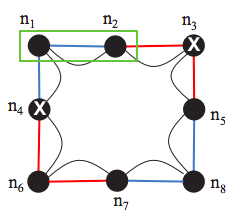} }}%
    		\caption{When the 2-qubit gate $b_{12}$, green box, is   executed, the neighboring qubits located at $n_3$ and $n_4$ can not be used by another gate. For instance, the gate $r_{35}$ can not be executed in parallel with $b_{12}$. In this case, the word $r_{35}b_{12}$ (and the word $b_{12}r_{35}$) can not be reduced. This is in contrast to the  absence of cross-talk, where  the rule
    		$r_{35}b_{12} \rightarrow rb_{35, 12}$  reduces
    		the word  $r_{35}b_{12}$ to the letter $rb_{35, 12}$ which has the weight of the maximum of the two weights ${\bf w}(r_{35})$ and ${\bf w}(b_{12})$. The rewriting system favors parallelism. }
           \label{crosstalk}
 	 \end{figure}

{\bf Parallelism.}
To favor parallelism,  we  enlarge our  set of letters to include additional letters that correspond to parallel gates.   As an example, consider the situation depicted in Figure \ref{crosstalk}. The fact that the two gates $b_{12}$ and $b_{58}$ can be run in parallel amounts to adding to $X^*$ the letter
$bb_{12; 58}$ and, to the system $\mathcal R$, the two rules
\begin{system}\label{commute}
 b_{12}b_{58} &\rightarrow& bb_{12; 58},\\
 b_{58}b_{12} &\rightarrow& bb_{12; 58}. 
\end{system}%
The new letter $ bb_{12, 58}$ weighs
the maximum of the two weights (durations)  ${\bf w}(b_{12})$ and  ${\bf w}(b_{58})$. We think of the letter $ bb_{12; 58}$ as a single task with ${\bf w}(bb_{12; 58}) = \mathrm{max}(\{{\bf w}(b_{12}),  {\bf w}(b_{58})\})$.

~~\\ 
{\bf Encoding cross-talk constraints. } In the presence of cross-talk constraints, which prevent gates
in physical proximity from being executed concurrently as in Figure \ref{crosstalk}, the procedure above is not applied to these ``cross-talking'' gates. An example is given by the two gates $b_{12}$ and $r_{35}$.  In which case, the two words $b_{12}r_{35}$ and $r_{35}b_{12}$ are kept without reduction, forcing  sequential execution. If, however, there is no cross-talk, 
 then similar rules as in (\ref{commute}) (as well as the letter $rb_{35,12}$) should be added, in order to favor parallelism, as mentioned.

\begin{figure}[H]
    	\centering
    		\subfloat  
    		{{\includegraphics[scale=0.6]{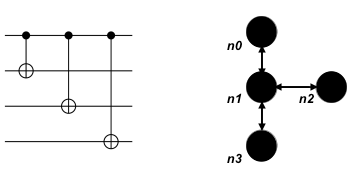} }}%
    		\caption{Initializations for the idealized circuit (left) on the hardware (right).
    		{ {}}
    		}
           \label{init}
 	 \end{figure}

~~\\
 {\bf Initial qubit placement.}
The initial placement of logical circuit qubits to hardware qubits can be read off from the reduction of the compiled gate. No extra work is needed specifically for this task. This is illustrated in the example of Figure \ref{init}, where the task is to compile the 2-gates $cnot_{01},$  $cnot_{02}$, and  $cnot_{03}$.

~~\\
If we place the $i$th circuit qubit at 
 $i$th hardware qubit $i=0,1,2,3$, our set of three gates is implemented as
 $cnot_{01} s_{01} cnot_{12} cnot_{13}$, which consumes one swap. Using the rule
 \begin{equation}
     cnot_{01} s_{01} \rightarrow cnot_{10},
 \end{equation}
we can reduce our initial implementation into 
 \begin{equation}
   cnot_{01} s_{01} cnot_{12} cnot_{13}\rightarrow cnot_{10} cnot_{12} cnot_{13}. 
 \end{equation}
 The new reduced form doesn't contain any swap, and the qubits placement can read from it when parsed from left to right: logical qubit 0 in hardware qubit 1, logical qubit 1 in hardware qubit location 0, and logical qubits 2 and 3 in hardware locations 2 and 3, respectively.

\section{Shuffle algebras: Return to commutativity} 
The {\it shuffle product}  \cite{Reutenauer} on  $\mathbb{Q}\langle X\rangle$ is defined (on $X^*$) recursively by:
\begin{system}
 \forall w\in X^*, && \varepsilon \,\sha\, w = 
 w \,\sha\, \varepsilon =w,\\[3mm]
 \forall x, y\in X, \, u,v\in X^*, &&
 xu \,\sha\, yv = x(u\,\sha\, yv) + y(xu\,\sha\, v).
\end{system}%
(and then linearly extended to the polynomial in $\mathbb{Q}\langle X\rangle$). With the shuffle product (replacing the simple concatenation),   $\mathbb{Q}\langle X\rangle$ is {\it commutative} (and continues to be {associative}) $\mathbb{Q}-$algebra, which we denote by $\mathrm{Sh}_\mathbb{Q}(X)$. It can be encoded elegantly using the Lyndon
words, which compresses expressions and makes calculations more efficient.

~~\\
Two words $u$ and $v$ are said to be conjugate if and only if 
\begin{equation}
    \exists x, y \in X^*\, \mbox{ such that } u=xy \mbox{ and } v=yx.
\end{equation}
This relation is an equivalence relation on $X^*$. As an example, consider the alphabet $X=\{a, b\}$. The words $a^2 b$,  $aba$ and $ba^2$ are all two by two conjugate. Intuitively, this consists of considering rotations (cyclic permutations) of letters. 

\begin{definition}
A word $w\in X^*$ is a Lyndon word if and only if it satisfies one of the two equivalent properties:
\begin{itemize}
    \item $w$ is  smaller than all of its rotations (conjugates).
    \item $w$ is  smaller than all of its right factors.
\end{itemize}
We write $L$ for the set of Lyndon words over $X$.
\end{definition}
A useful property that can be derived from this definition is that if $u$ and $v$ are two Lyndon words, then their  concatenation $uv$ is a Lyndon word if and only if $u\prec v$.  

\begin{proposition}[Standard factorization]
Each word $w$ in $X^*$ can be {\textit{uniquely}} written as 
\begin{equation}
    w = l_1 \cdots l_n
\end{equation}
where each $l_i$ is a Lyndon word with $l_1 \succ  \cdots \succ  l_n$.
\end{proposition}
This unique factorization is essential in our algorithm (compiler): although the shuffle product of two words is not a word, the final result of the compilation (if it exists) will uniquely be given in the form    
$w = l_1 \cdots l_n$--that is, a word! 
\begin{example}
The gates $w_1$ given by (\ref{w1}) 
factors into
\begin{equation}
  w_1 = [s_{14}s_{12}s_{23}s_{12}r_{23}];
\end{equation}
that is, $w_1$ is a Lyndon word. The word $w_3$, given by (\ref{w3}), factors into two
Lyndon words: 
\begin{equation}
   w_3 = [b_{12}][s_{14;23}s_{12}s_{23}], 
\end{equation} 
because $[b_{12}]\succ [s_{14;23}s_{12}s_{23}]$. 
\end{example}
 
\begin{proposition}[Basis of the shuffle algebra--Radford basis]
Let $L$ be the set of all Lyndon words over $X$. The algebra $\mathrm{Sh}_\mathbb{Q}(X)$ is a polynomial algebra that is isomorphic to $\mathbb{Q}[L]$.
\end{proposition}
Combining the last two propositions we have
\begin{proposition}
Any non-commutative polynomial in  $\mathbb{Q}\langle X\rangle$ can be
expressed in a unique way as a commutative polynomial on the Lyndon words. 
\end{proposition}
This means that one can use the Buchberger algorithm to compute the rewriting system $\mathcal R$ with the shuffle product replacing the usual product. In particular, the S--polynomial of $p$ and $q$ is the polynomial
\begin{equation}
 lt(p)\,\sha\, q - p\,\sha\, lt(q).    
\end{equation}
The subtlety, however, is how to express this as a rule mapping a word in $X^*$ (going back to concatenation) to a polynomial in $\mathbb Q\langle X\rangle$. The answer lies in the following ({\it triangular}) property:
\begin{proposition}
For each word $w$, written as a product of Lyndon words
$w=l_1^{i_1}\cdots l_k^{i_k}$ with $l_1\succ \cdots\succ l_k$ and $i_1,\cdots, i_k\geq 1$, one has
\begin{equation}
    \frac{1}{i_1!\cdots i_k!} l_1^{\sha\,i_1}\,\sha\,\cdots \,\sha\, l_k^{\sha\,i_k} = w +\sum_{u\prec w} \alpha_u u,
\end{equation}
for some natural integers $\alpha_u$.
\end{proposition}
\begin{example} Let us reproduce the same calculations (we have done using the Knuth-Bendix algorithm), now using the shuffle algebra.  
Consider again the two rules:
\begin{system}
 (s_{23} s_{12})^3&\rightarrow& \varepsilon,\\[3mm]
 s_{12}^2&\rightarrow& \varepsilon,
\end{system}%
which can be written in terms of the Lyndon words $l_1=s_{23} s_{12}$ and $l_2 =  s_{12}$ as  
\begin{system}\label{rulesshuffled}
  {l_1}^{\sha\, 3}&\rightarrow& \varepsilon,\\[3mm]
  {l_2}^{\sha\, 2}&\rightarrow& \varepsilon.
\end{system}%
On the other hand, from the definition of the shuffle algebra, we have $l_1 \sha\, l_2 \rightarrow l_3 \sha\, l_2 \sha\, l_2 $ with $l_3\ = s_{23}.$ This gives 
\begin{equation}
    l_1 \sha\, l_2 \rightarrow l_3.
\end{equation}
We replace this into the first equation of the system (\ref{rulesshuffled}), just after shuffling both sides of the equation with $l_2$. We obtain 
\begin{equation}
    l_1^{\sha\, 2} \sha\, l_3 \rightarrow l_2,
\end{equation}
which when expanded gives
\begin{equation}
    s_{23}s_{12}s_{23}s_{12}s_{23} \rightarrow s_{12}.
\end{equation}
This is the same rule (\ref{favouriteRule}) obtained using the Knuth-Bendix algorithm.
\end{example}

\section{Conclusion}
In this work, we have recast the problem of compiling quantum circuits in the language of rewriting systems. The rewriting system is deduced from the hardware constraints, either via Knuth-Bendix or through shuffle algebras.  This operation  needs to be done just once. Thus, compiling any input idealized gate is then simply computing its unique normal form with respect to the rewriting system. The normal form computation is a sequence of substitutions (rules). Future work should study these methods empirically.


\bibliographystyle{plain}
\bibliography{c1}
\end{document}